\documentclass[12pt]{iopart}
\usepackage{graphicx}
\usepackage{setstack}
\usepackage{iopams}

\newcommand{\dfrac}[2]{\frac{\displaystyle #1}{\displaystyle #2}}
\newcommand{\email}[1]{\ead{#1}}
\newcommand{\affiliation}[1]{\address{#1}}

\newcommand{\negthickspace}{\!}

\newcommand{\hypergauss}[4]{\phantom{}_{_2}\mathrm{F}\!_{_1}\!\left(#1,#2;#3;#4\right)}
\newcommand{\func}[2]{f_\mathrm{#1}\!\left(#2\right)}
\newcommand{\funcs}[1]{\func{}{#1}}
\newcommand{\funcc}[1]{\func{c}{#1}}

\newcommand{\Mpc}{\mathrm{Mpc}}
\newcommand{\Kpc}{\mathrm{kpc}}

\newcommand{\pc}{\mathrm{pc}}
\newcommand{\TeV}{\mathrm{TeV}}
\newcommand{\GeV}{\mathrm{GeV}}

\newcommand{\ud}{\mathrm{d}}
\newcommand{\uc}{\mathrm{c}}
\newcommand{\uh}{\mathrm{h}}

\newcommand{\uk}{\mathrm{k}}

\newcommand{\ui}{\mathrm{i}}
\newcommand{\ue}{\mathrm{e}}
\newcommand{\ueq}{\mathrm{eq}}
\newcommand{\umat}{\mathrm{mat}}
\newcommand{\urad}{\mathrm{rad}}
\newcommand{\uini}{\mathrm{ini}}
\newcommand{\ur}{\mathrm{r}}
\newcommand{\um}{\mathrm{m}}
\newcommand{\uM}{\mathrm{M}}
\newcommand{\uR}{\mathrm{R}}

\newcommand{\calS}{\mathcal{S}}
\newcommand{\calP}{\mathcal{P}}
\newcommand{\calF}{\mathcal{F}}
\newcommand{\calN}{\mathcal{N}}
\newcommand{\calI}{\mathcal{I}}
\newcommand{\calIc}{\mathcal{I}_\uc}
\newcommand{\calNini}{\calN_\uini}
\newcommand{\xini}{x_\uini}

\newcommand{\mpl}{m_{_{\mathrm Pl}}}

\newcommand{\const}{C_\circ}
\newcommand{\Konst}{K}

\newcommand{\scaling}{\calS}

\newcommand{\horizon}{d_\uh}

\newcommand{\rhoinf}{\rho_\infty}
\newcommand{\rholoop}{\rho_\circ}
\newcommand{\chic}{\chi_\uc}
\newcommand{\chimat}{\chi_{_\uM}}
\newcommand{\chirad}{\chi_{_\uR}}
\newcommand{\Crad}{C_{_\uR}}
\newcommand{\Cmat}{C_{_\uM}}
\newcommand{\Cc}{C_\uc}
\newcommand{\Ci}{C_\ui}

\newcommand{\OmegaR}{\Omega_{\ur_0}}
\newcommand{\OmegaM}{\Omega_{\um_0}}
\newcommand{\OmegaO}{\Omega_\circ}
\newcommand{\gammad}{\gamma_\ud}
\newcommand{\gammac}{\gamma_\uc}
\newcommand{\gammax}{\gamma_\tau}
\newcommand{\gammai}{{\gamma_\infty}}
\newcommand{\muc}{\mu_\uc}

\newcommand{\zeq}{z_\ueq}
\newcommand{\zcross}{z_{*}}
\newcommand{\zini}{z_\uini}
\newcommand{\zh}{z_\uh}
\newcommand{\zd}{z_\ud}
\newcommand{\zp}{z_{_\%}}
\newcommand{\tini}{t_\uini}
\newcommand{\tx}{t_\tau}
\newcommand{\thor}{t_\uh}
\newcommand{\tcross}{t_*}
\newcommand{\aini}{a_\uini}

\begin{document}

\title{Cosmic string loop distribution on all length scales 
and at any redshift}

\author{Larissa Lorenz}
\email{larissa.lorenz@uclouvain.be}
\affiliation{Institute of Mathematics and Physics, Centre for
  Cosmology, Particle Physics and Phenomenology, \\ Louvain
  University, 2 Chemin du Cyclotron, 1348 Louvain-la-Neuve, Belgium}

\author{Christophe Ringeval}
\email{christophe.ringeval@uclouvain.be}
\affiliation{Institute of Mathematics and Physics, Centre for
  Cosmology, Particle Physics and Phenomenology, \\ Louvain
  University, 2 Chemin du Cyclotron, 1348 Louvain-la-Neuve, Belgium}

\author{Mairi Sakellariadou}
\email{mairi.sakellariadou@kcl.ac.uk}
\affiliation{Department of Physics, King's College, University of
London, Strand, London WC2R 2LS, United Kingdom}

\date{\today}

\begin{abstract}
  We analytically derive the expected number density distribution of
  Nambu--Goto cosmic string loops at any redshift soon after the time
  of string formation to today. Our approach is based on the
  Polchinski--Rocha model of loop formation from long strings which we
  adjust to fit numerical simulations and complement by a
  phenomenological modelling of gravitational
  backreaction. Cosmological evolution drives the loop distribution
  towards scaling on all length scales in both the radiation and
  matter era. Memory of any reasonable initial loop distribution in
  the radiation era is shown to be erased well before Big Bang
  Nucleosynthesis. In the matter era, the loop distribution reaches
  full scaling, up to some residual loops from the radiation era which
  may be present for extremely low string tension. Finally, the number
  density of loops below the gravitational cutoff is shown to be scale
  independent, proportional to a negative power of the string tension
  and insensitive to the details of the backreaction modelling. As an
  application, we show that the energy density parameter of loops
  today cannot exceed $10^{-5}$ for currently allowed string tension
  values, while the loop number density cannot be less than $10^{-6}$
  per $\Mpc^{3}$. Our result should provide a more robust basis for
  studying the cosmological consequences of cosmic string loops.
\end{abstract}

\maketitle

\section{Introduction}
\label{sec:intro}

It was shown by Kibble that cosmic strings are a natural outcome of
cosmological phase transitions~\cite{Kibble:1976, Hindmarsh:1994re,
  Vilenkin:2000, Sakellariadou:2006qs}, and their possible presence in
our Universe has been extensively studied ever since. Cosmic strings
come in the form of line-like topological defects produced by the
spontaneous breakdown of some symmetry~\cite{Kirzhnits:1972,
  Jeannerot:2003qv, Rocher:2004my}, but can also be stretched
superstrings generated at the end of brane
inflation~\cite{Burgess:2001fx, Sarangi:2002yt, Dvali:2003zj,
  Jones:2003da, Davis:2005, Sakellariadou:2008ie, Copeland:2009ga,
  Sakellariadou:2009ev}. Once formed, cosmic strings evolve and very
rapidly reach their so-called scaling regime in which the energy
density $\rhoinf$ of super-horizon sized strings behaves as radiation
in the radiation era, or matter in the matter era~\cite{Albrecht:1989,
  Bennett:1989, Bennett:1990, Allen:1990}. Hence, for these long
strings it holds that $\rhoinf \propto 1/\horizon^2$, where $\horizon$
stands for the distance to the horizon ($\horizon \propto t$, the
cosmic time). This scaling ensures that long strings never dominate
the energy density of the Universe at late times and remain compatible
with cosmological observations provided their energy density per unit
length $U$ is small enough. The current Cosmic Microwave Background
(CMB) data limits the string contribution to at most $10\%$ on the
angular scales observed by the Wilkinson Microwave Anisotropies Probe
(WMAP) satellite~\cite{Jarosik:2010iu}, i.e. $GU < 7\times 10^{-7}$
for Abelian vortices, $G$ being the Newton
constant~\cite{Bevis:2007gh}. However, being present all along the
Universe's history, cosmic strings may be the dominant source of CMB
non-gaussianities and are expected to dominate over inflationary
perturbations at small angular scales~\cite{Fraisse:2007nu,
  Takahashi:2008ui, Hindmarsh:2009qk, Hindmarsh:2009es, Regan:2009hv,
  Yamauchi:2010vy, Landriau:2010cb, Bevis:2010gj, Ringeval:2010ca}.

The existence of a scaling regime for Nambu--Goto (NG) string networks
requires the incessant formation of loops from long cosmic
strings. Loops are produced at string autocommutation and intersection
events, thereby evacuating the excess energy density from long
strings. In the standard picture, the produced loops shrink and
disappear by gravitational wave emission, which could provide
additional observational evidence for cosmic
strings~\cite{PhysRevD.45.1898, Damour:2001bk, Siemens:2006vk,
  Olmez:2010bi, Dubath:2007wu}. Note, however, that Abelian Higgs
string networks preferentially reach scaling by boson radiation rather
than loop formation, at least during their numerically probed
evolution~\cite{Vincent:1998, Moore:2002, Hindmarsh:2008dw}.
  
The expected cosmological distribution of loops has been subject to
intense debate since the development of the original ``one scale''
model by Kibble~\cite{Kibble:1976}.  In this model, at any time loops
are formed with a size given by the typical distance separating long
strings, i.e. a fraction of the distance to the horizon $\horizon$.
NG numerical simulations in
Friedmann--Lema\^{\i}tre--Robertson--Walker (FLRW) spacetime have
shown, however, that most loops are formed at much smaller size,
suggesting the existence of tiny correlation lengths in cosmic string
networks~\cite{Bennett:1989, Allen:1990, Vincent:1996rb}. These
simulations are nevertheless limited in time and cannot exceed a
redshift range typically of order $10^2$. Comparing this to the
$10^{18}$ change in redshift expected at Big Bang Nucleosynthesis
(BBN) [provided strings are formed at the Grand Unified Theory (GUT)
energy scale], it is crucial for simulations to disentangle transient
effects from the ones lasting over cosmological timescales.

As shown in Refs.~\cite{Vanchurin:2005yb, Ringeval:2005kr}, many of
the small loops observed in Nambu--Goto simulations indeed come from a
transient effect associated with the relaxation of the initial string
network towards its stable configuration. One must not expect these
loops to remain present after a long enough evolution. However, in
addition to transient loops, Ref.~\cite{Ringeval:2005kr} was the first
to exhibit a sub-population of loops in scaling which constitute the
cosmological attractor, a result recovered in
Refs.~\cite{Vanchurin:2005pa, Martins:2005es, Olum:2006ix}, albeit
with some differences in the distribution. The energy density
distribution of loops in scaling follows an evolution similar to long
strings, i.e. it behaves as radiation in the radiation era, and as
matter in the matter era. In this regime, the loop number density
distribution has a power law shape: it is ``scale-free'' in the sense
that loops of all size are present at any time. These results can be
explained if the dominant mechanism of string evolution also applies
to loops: they mainly loose energy by in turn emitting other
loops. More quantitatively, denoting by $\alpha=\ell/\horizon$ the
size of a loop in units of $\horizon$, the loop number density
$n(\alpha,t)$ and energy density $\rholoop(\alpha,t)$ in scaling
read~\cite{Ringeval:2005kr}
\begin{equation}
\label{eq:scaling}
\dfrac{\ud n}{\ud
  \alpha} = \dfrac{\scaling(\alpha)}{\alpha \, \horizon^3}\,, \qquad \dfrac{\ud \rholoop}{\ud \alpha} = U \dfrac{\scaling(\alpha)}{\horizon^2}\,,
\end{equation}
where the ``scaling function'' $\scaling(\alpha)$ is well fitted, in
the radiation and matter era, by the power laws $\scaling(\alpha) =
\const \, \alpha^{-p}$ with~\cite{Ringeval:2005kr}
\begin{equation}
\label{eq:power-law}
\begin{array}{cc}
\left\{
\begin{array}{ccc}
p &= & 1.41\,^{+0.08}_{-0.07} \\
\const & = & 0.09\,^{-0.03}_{+0.03}
\end{array}
\right|_\umat,
\quad
\left\{
\begin{array}{ccc}
p & = & 1.60\,^{+0.21}_{-0.15} \\
\const & = & 0.21\,^{-0.12}_{+0.13}
\end{array}
\right|_\urad.
\end{array}
\end{equation}
Note that the energy density of loops evolves as does 
the one of long strings, $\rholoop\propto1/\horizon^{2}$.

NG numerical simulations only account for string dynamics and
intercommutation events, and in particular they know nothing about the
impact of the loops' gravitational wave emission. In a realistic
physical situation, a loop of size $\ell$ emitting gravitational waves
is expected to decay within a time roughly given by $t_\ud =
\ell/(\Gamma GU)$, where $\Gamma$ is a numerical coefficient depending
on the string structure, which for typical loops is of the order of
$\Gamma \simeq 10^2$~\cite{Vilenkin:1981, PhysRevD.45.1898}. As a
result, for small loops with $\alpha < \alpha_\ud \propto \Gamma G U$,
gravitational wave emission should overcome loop emission as the
dominant energy loss mechanism. The (numerically found) scaling law of
Eq.~(\ref{eq:power-law}) can therefore no longer be used in the regime
$\alpha < \alpha_\ud$, and the loop distribution at those length
scales must be understood further analytically, taking into account
the emission of gravitational waves.

Another phenomenon associated with gravitational radiation is the
so-called backreaction effect which applies to both loops and long
strings. The emission of gravitational waves renders the strings
smoother and smoother on the smallest length scales, hence preventing
self-intersection over a very small
distance~\cite{Sakellariadou:1990ne, Hindmarsh:1990xi}. Therefore,
long strings and loops cannot produce infinitely small loops:
gravitational backreaction cuts off loop production below a certain
scale. The typical length at which this happens was estimated to be of
order $\ell_\uc \propto \left(GU\right)^{1+2\chi} t$, where
$\chi>0$. When translated into the variable $\alpha$ and compared to
the scale of gravitational decay $\alpha_\ud \propto \Gamma G U$
introduced above, this corresponds to a constant value of $\alpha_\uc
< \alpha_\ud$~\cite{Siemens:2001dx, Siemens:2002dj,
  Polchinski:2007rg}.

The goal of this paper is to establish, both in the radiation and the
matter dominated eras, a phenomenological description of the string
loop number density evolution which includes the three above-mentioned
effects: loop formation, decay by gravitational wave emission and the
impact of gravitational backreaction. Various analytical models have
been proposed so far to describe the evolution of a cosmic string
network~\cite{Austin:1993rg, Martins:1996jp, Copeland:1998na,
  Martins:2000cs}. Each of these models starts from a different set of
assumptions and consequently, their results may differ. Our objective
here is to describe the cosmic string loop distribution in the
well-motivated situation where it coexists with a long string network
in its scaling regime. Long strings indeed reach their cosmological
attractor after a redshift range which does not exceed an order of
magnitude~\cite{Ringeval:2010ca}. For this reason, we use the
Polchinski--Rocha (PR) approach of Refs.~\cite{Polchinski:2006ee,
  Dubath:2007mf} in which loops are formed from tangent vector
correlations along long strings in scaling. Let us mention that a
non-scaling loop production function has recently been introduced in
Refs.~\cite{Vanchurin:2010me, Vanchurin:2010tk} to explain the
existence of loops associated with the initial conditions. As
discussed below (see Sec.~\ref{sec:radera}), our approach accommodates
the existence of such loops without invoking a non-scaling loop
production function.

The PR model predictions are in good agreement with the correlators
measured in Abelian Higgs simulations~\cite{Hindmarsh:2008dw} and also
with the NG loop distribution of
Eq.~(\ref{eq:power-law})~\cite{Rocha:2007ni}, up to the overall
normalisation. This agreement may appear surprising since simulations
include many additional effects on top of loop formation from long
strings, e.g. loop formation from loops, loop fragmentation and loop
reconnection leading to daughter loops of a different size than the
parent ones, all of them being non-negligible~\cite{Bennett:1990}. The
physical interpretation is that these additional mechanisms preserve
the functional form of the analytical PR loop production function
(which is derived for the long strings) and amount to a
renormalization of its amplitude. The PR production function can thus
be considered as a robust physical feature against these effects, and
for this reason, we take it as an ansatz for the \emph{total} loop
production function yet to be established. However, we can no longer
fix its unknown coefficients analytically by using the long string
scaling properties since we want to extend its use beyond the ``long
strings only'' picture.  We can, however, choose these coefficients to
match the numerical results of Eq.~(\ref{eq:power-law}), in the
appropriate range of length scales.

Below, we extend the method originally introduced by Rocha in
Ref.~\cite{Rocha:2007ni} along the three following
directions. Firstly, we introduce a change in behaviour of the loop
production function below the gravitational backreaction length
scale. The true shape of the loop production function in this regime
being unknown, we consider various motivated possibilities and show
that the final loop number density distribution remains
unchanged. Secondly, we do not \emph{a priori} neglect the transient
solutions, which allows us to discuss the effect of the initial loop
number density distribution at string formation time. This is the
subject of Sec.~\ref{sec:evoleqs}. In Sec.~\ref{sec:cosmo}, we discuss
how the distribution evolves with redshift through the radiation and
matter eras, as well as through the transition. We find that all loops
rapidly reach the so-called ``full-scaling regime'', in which their
number density assumes the universal form plotted in
Fig.~\ref{fig:scalrad}. We furthermore derive the loops' energy
density parameter and the loop number density today before presenting
our conclusions in Sec.~\ref{sec:conclusions}.

\section{Cosmological attractor}
\label{sec:evoleqs}
Borrowing notation from Ref.~\cite{Rocha:2007ni}, the distance to the
horizon in terms of cosmic time $t$ is expressed as
\begin{equation}
\label{eq:d_h}
\horizon(t) = \dfrac{t}{1-\nu}\,,
\end{equation}
where the scale factor $a(t) \propto t^\nu$, with $\nu=1/2$ in the
radiation era and $\nu=2/3$ in the matter era. Moreover, the cosmic
time $t$ can be expressed in terms of redshift $z$, of the Hubble
parameter and of the density parameters today ($H_{0}$, $\OmegaR$ and
$\OmegaM$, respectively) as
\begin{equation}
\label{eq:trad}
  t_\urad(z)  \simeq \dfrac{1}{2 H_0 \sqrt{\OmegaR}} \dfrac{1}{(1+z)^2}\,,
\end{equation}
in the radiation era and as
\begin{equation}
\label{eq:tmat}
  t_\umat(z) \simeq \dfrac{2}{3 H_0 \sqrt{\OmegaM}} \dfrac{1}{(1+z)^{3/2}}\,,
\end{equation}
in the matter era. The above expressions constitute a very good
approximation of the currently favoured $\Lambda$CDM (Lambda Cold Dark
Matter) model~\cite{Komatsu:2010fb} up to the smooth transition from
the radiation to the matter era, and before the cosmological constant
domination sets in (i.e. for redshifts $z > 2$). Note that, from
Eqs.~(\ref{eq:trad}) and (\ref{eq:tmat}), the instantaneous transition
between the two eras occurs at
\begin{equation}\label{eq:zcross}
  \zcross = \dfrac{9}{16} \dfrac{\OmegaM}{\OmegaR} -1 
  = \dfrac{9}{16}(1 + \zeq) - 1,
\end{equation}
where $\zeq$ stands for the redshift of equality between the radiation
and matter energy densities.

\subsection{Evolution equation}

Denoting by $n(\ell,t)$ the number density distribution of cosmic
string loops of size $\ell$ at cosmic time $t$, we can write in
the Eulerian description
\begin{equation}
\label{eq:dpart}
\dfrac{\ud}{\ud t} \left(a^3 \dfrac{\ud n}{\ud \ell} \right) 
= a^3 \calP(\ell,t),
\end{equation}
where $\calP(\ell,t)$ is the loop production function, i.e. the number
density distribution of loops of size $\ell$ produced per unit of time
at $t$. In the Lagrangian description, due to gravitational radiation,
a loop of size $\ell(t)$ shrinks at a constant rate such that
\begin{equation}
\label{eq:gammad}
\dfrac{\ud \ell}{\ud t} = -\gammad,
\end{equation}
which we take as the definition of $\gammad$. According to the
discussion of Sec.~\ref{sec:intro}, and using Eqs.~(\ref{eq:d_h}) and
(\ref{eq:gammad}), we have
\begin{equation}
\label{eq:gammagu}
  \gammad = \dfrac{\alpha_\ud}{1-\nu}\simeq \Gamma GU.
\end{equation}
Combining Eqs.~(\ref{eq:dpart}) and (\ref{eq:gammad}), one
gets~\cite{Rocha:2007ni}
\begin{equation}
\label{eq:evolell}
  \dfrac{\partial}{\partial t} \left(a^3 \dfrac{\ud n}{\ud \ell}
  \right) - \gammad \dfrac{\partial}{\partial \ell}\left(a^3
  \dfrac{\ud n}{\ud \ell} \right) = a^3 \calP(\ell,t).
\end{equation}
As suggested by the form of the scaling distribution in
Eq.~(\ref{eq:scaling}), it is more convenient to express the loop
sizes $\ell$ in units of the horizon $\horizon$. In fact, since we shall be
interested in the radiation matter transition, it is even more
convenient to work in units of cosmic time $t$. For this purpose,
let us define a new variable $\gamma$ and a new function $\calF$, as
\begin{equation}
  \gamma(\ell,t) \equiv \dfrac{\ell}{t}\,, \qquad \calF(\gamma,t)
  \equiv \dfrac{\ud n}{\ud \ell}\,,
\end{equation}
respectively, in terms of which Eq.~(\ref{eq:evolell}) reads
\begin{equation}
\label{eq:evolgam}
t \dfrac{\partial(a^3 \calF)}{\partial t} - \left(\gamma + \gammad
\right) \dfrac{\partial(a^3 \calF)}{\partial \gamma} = a^3 t
\calP(\gamma,t).
\end{equation}
Evidently, the time variable $t$ ranges from the initial time of 
string network formation ($t_\uini$) to today ($t_{0}$), while 
$0\leq\gamma\leq\gamma_{\mathrm{max}}$, where 
$\gamma_{\mathrm{max}}$ corresponds to 
$l_{\mathrm{max}}$, the size of the 
largest (horizon-sized) loops, which will be discussed below.

\subsection{Loop production function}

\begin{figure}
\begin{center}
\includegraphics[width=0.65\textwidth]{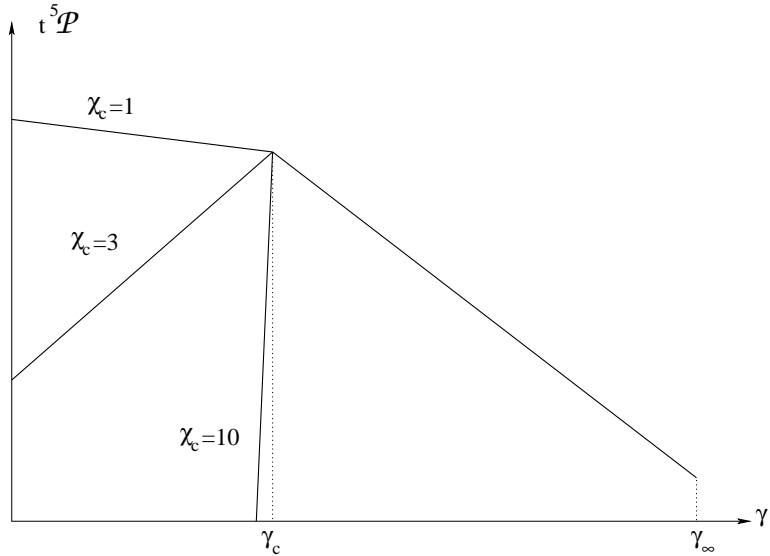}
\caption{Sketch of the loop production function $t^5
  \calP(\gamma,t)$ as a function of $\gamma$ in logarithmic units. The
  gravitational backreaction scale is located at $\gamma = \gammac$.}
\label{fig:loopprod}
\end{center}
\end{figure}

The scaling law of Eq.~(\ref{eq:scaling}) implies that, once the
scaling regime is reached, $t^4 \calF(\gamma,t)$ should be a function
of $\gamma$ only. From Eq.~(\ref{eq:evolgam}), we expect the same to
happen for $t^5 \calP(\gamma,t)$. Following
Refs.~\cite{Polchinski:2006ee, Dubath:2007mf, Rocha:2007ni}, we
moreover assume that this function is a power law, namely
\begin{equation}
\label{eq:loopprod}
t^5 \calP(\gamma,t) =c\,\gamma^{2\chi-3}\,,
\end{equation}
where $c$ and $\chi$ are two parameters that will be fixed to fit
Eq.~(\ref{eq:power-law}). However, according to our discussion of
Sec.~\ref{sec:intro}, the above expression is only valid for a range
of $\gamma$ values where gravitational backreaction effects can be
neglected. Hence, Eq.~(\ref{eq:loopprod}) holds for values of $\gamma$
greater than
\begin{equation}
\label{eq:gammac}
  \gammac\equiv \dfrac{\alpha_\uc}{1-\nu} \simeq \Upsilon (GU)^{1+2\chi},
\end{equation}
where $\Upsilon$ is a number
$\mathcal{O}(10)$~\cite{Polchinski:2007rg}.  The parameter $\chi$ in
this expression is expected to be the same as in
Eq.~(\ref{eq:loopprod}), given that the gravitational backreaction
length scale is precisely derived from the same two-point correlators
used in the PR model to obtain Eq.~(\ref{eq:loopprod}) (see
Ref.~\cite{Polchinski:2007rg}). In our phenomenological approach, we
consider $\gammac$ as a free parameter below which the loop production
function is given by a different power law,
\begin{equation}
\label{eq:tinyprod}
  t^5 \calP(\gamma<\gammac,t) = c_\uc\,\gamma^{2\chi_\uc - 3}\,.
\end{equation}
The parameters $c_\uc$ and $\chi_\uc$ are such that the loop
production function is continuous across $\gamma=\gammac$, 
which leads to
\begin{equation}
  \label{eq:crels}
  c_\uc = c \, \gammac^{2(\chi - \chic)} .
\end{equation}
Moreover, since gravitational backreaction smooths the 
strings on the smallest length scales, less loops should be produced in
the small $\gamma$ limit, and one may naively expect the 
production function $\calP(\gamma,t)$ to vanish for
$\gamma=0$. In fact, by incorporating backreaction effects into
the long string structure, it has been shown in
Ref.~\cite{Polchinski:2007rg} that, for $\gamma \simeq \gammac$,
Eq.~(\ref{eq:loopprod}) has a power law exponent renormalised to
$\chi_\mathrm{br} \simeq 1$. This gives an extreme lower limit
for $\chi_\uc$, 
\begin{equation}
\chic \ge 1\,,
\end{equation}
as backreaction cannot smooth the string structure less than
that. From Eq.~(\ref{eq:tinyprod}) we therefore see that the loop
production function can be divergent in the $\gamma \rightarrow 0$
limit for $1 <\chic < 3/2$. However, being its first integral, the loop
number density $\calF(\gamma,t)$ will remain finite. Only in the
limiting case $\chic=1$ would we find a logarithmic
$\gamma\rightarrow0$ divergence for $\calF(\gamma,t)$ (see below),
which nevertheless leads to a finite energy density (which is $\propto
U\ell \calF$).  This extreme situation corresponds to the production
of an infinite number of infinitely small loops but with finite energy
density. On the opposite, if gravitational backreaction was infinitely
effective, no loops at all would be produced with $\gamma<\gammac$,
corresponding to the limit $\chic \rightarrow \infty$. Certainly the
true physical situation lies somewhere between these two
extremes~\cite{Austin:1993rg}. In the following, we show that the
final loop number density distribution is (almost) insensitive to the
chosen value of $\chic$, and thus to the details of the backreaction
smoothing process. Only the cutoff scale $\gammac$ enters into the
expression for $t^{4}\calF(\gamma,t)$ in its scaling regime.  A sketch
of the loop production function $t^5 \calP(\gamma,t)$ as a function of
$\gamma$ is drawn in Fig.~\ref{fig:loopprod}.

With Eqs.~(\ref{eq:loopprod}) and (\ref{eq:tinyprod}) we now have at
our disposal a continuous expression for the loop production function
for all $\gamma\geq0$. As mentioned earlier, however, there exists an
upper bound $\gamma_{\mathrm{max}}$ on the accessible range of
$\gamma$: any loop with $\gamma > 1/(1-\nu)$ is of super-horizon size
and hence indistinguishable from a long string. Hence, its
contribution has already been included in the loop production
function. In fact, in a cosmological string network, there does exist
a small population of loops with size comparable to $\horizon$.  We
shall refer to them as ``Kibble's loops'', and they are not described
by our current approach~\cite{Ringeval:2005kr, Vanchurin:2005pa}. Due to
Hubble damping, their properties are in all aspects comparable to
those of long strings, and they essentially decay by loop emission
rather than gravitational radiation. As for long strings, we consider
their net effect included in the loop production function. The typical
size of these loops is given by the distance between two super-horizon
sized strings in scaling, which is
\begin{equation}
  \gammai = \dfrac{1}{1-\nu} \left(\dfrac{U}{\rho_\infty \horizon^2}
  \right)^{1/2}.
\end{equation}
From the values of $\rho_\infty$ given in Ref.~\cite{Ringeval:2005kr},
one gets $\gammai \simeq 0.32$ in the radiation era and $\gammai
\simeq 0.56 $ in the matter era. As we shall see later, our model
predicts no loop larger than these values. However, in a
realistic situation one should bare in mind that typically a few 
Kibble's loops are present.

Finally, it has been shown in Ref.~\cite{Copeland:2009dk} that kinks
propagating along the strings can also smooth the tangent vector
correlator used in the PR model of loop formation. This amounts to
setting $\chi=1/2$ in Eq.~(\ref{eq:loopprod}) for the $\gamma$ values
at which this effect appears, i.e. it would require the addition of
another piecewise domain in the definition of the loop production
function. Although possible, we have chosen not to include kink
smoothing in the following considerations since this effect is
expected to be significant in a domain $\gamma < \gamma_\uk(t)$, where
$\gamma_\uk(t)$ is a decreasing function of
time~\cite{Copeland:2009dk}. As a result, the effect is transient and
should no longer be observable when $\gamma_\uk(t)$ hits the
gravitational backreaction length scale $\gammac$.

\subsection{Evolution from an initial distribution}

We now assume the loop number density distribution is known at
some initial time $\tini$, when it is given by
\begin{equation}
\calNini(\ell) \equiv \dfrac{\ud n}{\ud \ell} (\ell,\tini).
\end{equation}
For instance, the initial loop distribution can be obtained by
examining the statistical properties of the Higgs field phases just
after the phase transition responsible for string
formation~\cite{Vachaspati:1984dz, Karra:1997it,
  Hindmarsh:2001vp}. However, since the loop production function of
Eq.~(\ref{eq:loopprod}) assumes a scaling form, $\calNini(\ell)$
should be understood as the loop distribution expected at least just
after the scaling of long strings takes place. In the following, we
solve Eq.~(\ref{eq:evolgam}), with the two-fold loop production
function of Eqs.~(\ref{eq:loopprod}) and (\ref{eq:tinyprod}), starting
from a given $\calNini$ at $t=\tini$.

Using new variables $w$ and $s$ defined from
\begin{equation}
w = \ln t, \qquad s = \ln t + \ln(\gamma + \gammad),
\end{equation}
as well as Eq.~(\ref{eq:loopprod}) for the loop production function,
Eq.~(\ref{eq:evolgam}) simplifies to
\begin{equation}\label{eq:evolws}
  \dfrac{\partial}{\partial w}\left(a^3 \calF \right) = c \, a^3\ue^{-4 w}
  \left(\ue^{s-w} - \gammad \right)^{2 \chi-3},
\end{equation}
for $\gamma > \gammac$. The same expression holds for $\gamma<\gammac$
up to the replacement $c\rightarrow c_\uc$ and $\chi \rightarrow
\chic$. Assuming $\gammad$ constant and $a \propto t^\nu$, this
equation is readily integrated in terms of the Gauss hypergeometric
function. For $\gamma> \gammac$, the solution of Eq.~(\ref{eq:evolws}) reads
\begin{equation}
 \calF(\gamma,t)  = \dfrac{c}{\mu} \dfrac{ \left(\gamma + \gammad
    \right)^{2\chi-3}} {t^4} \hypergauss{3 - 2
    \chi}{\mu}{\mu+1}{\dfrac{\gammad}{\gamma + \gammad}}  +
  \dfrac{\calI(\gamma t + \gammad t)}{a^3},
\end{equation}
where $\calI(x)$ is an arbitrary function (still to be
determined) and
\begin{equation}\label{eq:defmu}
\mu \equiv 3 \nu - 2\chi - 1.
\end{equation}
For $\gamma< \gammac$, one obtains exactly the same expression up to
the replacement of $c$ and $\chi$ into $c_\uc$ and $\chic$ [hence, $\muc$ is 
defined as in Eq.~(\ref{eq:defmu}) by replacing $\chi\rightarrow\chic$]. The
unknown integration function in this regime is denoted by $\calIc(x)$. 
In order to simplify the notations, we also define
\begin{equation}\label{eq:deffunc}
\funcs{x}  \equiv \hypergauss{3-2\chi}{\mu}{\mu+1}{x}, \quad
\funcc{x}  \equiv \hypergauss{3-2\chic}{\muc}{\muc+1}{x},
\end{equation}
and the rescaled constants
\begin{equation}
  \label{eq:constsc}
  C  \equiv \dfrac{c}{\mu}\,, \qquad \Cc \equiv \dfrac{c_\uc}{\muc}
  = \dfrac{\mu}{\muc} \gammac^{2(\chi - \chic)} C\,.
\end{equation}
The loop number density distribution in its two respective domains
then reads
\begin{eqnarray}
  \label{eq:solgama}
  \calF(\gamma \ge \gammac,t) & = & C \dfrac{\left (\gamma + \gammad
    \right)^{2\chi-3}}{t^4}  \funcs{\dfrac{\gammad}{\gamma + \gammad}}
  + \dfrac{\calI[(\gamma + \gammad)t]}{a^3}\,, \\
\calF(\gamma<\gammac,t) & = &\Cc \dfrac{\left (\gamma + \gammad
    \right)^{2\chic-3}}{t^4}  \funcc{\dfrac{\gammad}{\gamma + \gammad}}
   + \dfrac{\calIc[(\gamma + \gammad)t]}{a^3}\,.
\label{eq:solgamb}
\end{eqnarray}
Continuity of $\calF(\gamma,t)$ across $\gamma=\gammac$ at all times
imposes that the unknown functions $\calI(x)$ and $\calIc(x)$ are not
independent. From Eqs.~(\ref{eq:solgama}) and (\ref{eq:solgamb}) one
then obtains the relation
\begin{equation}
\label{eq:funcrel}
  \calIc(x) = \calI(x) + \Konst\dfrac{\left(\gammad +
    \gammac\right)^4}{x^4} \, \left[a \negthickspace
    \left(\dfrac{x}{\gammac + \gammad}\right)\right]^3,
\end{equation}
where $\Konst$ is a constant given by
\begin{eqnarray}
\label{eq:konst}
  \Konst = C (\gammac + \gammad)^{2\chi-3} \funcs{ \dfrac{\gammad}
    {\gammac + \gammad}} - \Cc (\gammac+\gammad)^{2\chic-3}
  \funcc{\dfrac{\gammad}{\gammac+\gammad}} .
\end{eqnarray}
Given $\calNini(\ell)$, we can now uniquely determine the two unknown
functions. At $t=\tini$, defining
\begin{equation}
\xini \equiv (\gammac+\gammad) \tini \, ,
\end{equation}
one can use Eq.~(\ref{eq:solgama}) to get $\calI(x)$ for all $x>\xini$:
\begin{equation}
\label{eq:funcs}
  \calI(x)  = \aini^3\, \calNini \negthickspace \left(x - \gammad
  \tini\right) - C \dfrac{\aini^3}{\tini^4}
  \left(\dfrac{x}{\tini} \right)^{2 \chi-3} \funcs{\dfrac{\gammad}{x}
    \tini}.
\end{equation}
This expression also provides $\calIc(x)$ for all $x>\xini$ by using
Eq.~(\ref{eq:funcrel}). Similarly, setting $t=\tini$ into
Eq.~(\ref{eq:solgamb}) uniquely determines $\calIc(x)$ for all
$x<\xini$,
\begin{equation}
\label{eq:funcc}
  \calIc(x) = \aini^3\, \calNini \negthickspace \left(x - \gammad
    \tini \right) - \Cc \dfrac{\aini^3}{\tini^4}
  \left(\dfrac{x}{\tini} \right)^{2 \chic-3} \funcc{\dfrac{\gammad}{x}
    \tini},
\end{equation}
which now gives $\calI(x<\xini)$ from Eq.~(\ref{eq:funcrel}). After
some algebra, plugging Eqs.~(\ref{eq:funcrel}), (\ref{eq:funcs}) and
(\ref{eq:funcc}) into Eq.~(\ref{eq:solgama}) and (\ref{eq:solgamb}),
where the argument of the function $\calI$ is now given by
$(\gamma+\gammad)t$, yields the final solution
\begin{eqnarray}
\fl t^4 \calF(\gamma \ge \gammac,t)  &=&  \left(\dfrac{t}{\tini} \right)^4
\left(\dfrac{\aini}{a}\right)^3 \tini^4\,
\calNini\negthickspace\left\{\left[\gamma + \gammad\left(1 -
      \dfrac{\tini}{t} \right) \right] t \right\} \nonumber \\   && + C(\gamma +
\gammad)^{2\chi-3} \funcs{\dfrac{\gammad}{\gamma + \gammad}}  \nonumber \\  && -
C (\gamma + \gammad)^{2\chi - 3} \left(\dfrac{t}{\tini}
\right)^{2\chi+1} \left(\dfrac{\aini}{a} \right)^3
\funcs{\dfrac{\gammad}{\gamma + \gammad} \dfrac{\tini}{t}}, \nonumber \\
\fl t^4\calF(\gammax \le \gamma < \gammac,t)  &=&
\left(\dfrac{t}{\tini} \right)^4 \left(\dfrac{\aini}{a}\right)^3
\tini^4\, \calNini\negthickspace\left\{\left[\gamma + \gammad\left(1
      - \dfrac{\tini}{t} \right) \right] t \right\} \nonumber \\ && +  \Cc (\gamma +
\gammad)^{2 \chic - 3} \funcc{\dfrac{\gammad}{\gamma + \gammad}} \nonumber \\ && - C
(\gamma + \gammad)^{2\chi - 3} \left(\dfrac{t}{\tini}
\right)^{2\chi+1} \left(\dfrac{\aini}{a} \right)^3
\funcs{\dfrac{\gammad}{\gamma + \gammad} \dfrac{\tini}{t}} \nonumber \\ && + \Konst
\left(\dfrac{\gammac + \gammad}{\gamma + \gammad} \right)^4 \,
\left[\dfrac{a \negthickspace \left(\dfrac{\gamma+\gammad}
      {\gammac+\gammad}t\right)} {a(t)} \right]^3 \, , \nonumber \\
\label{eq:finalsol}
\fl t^4 \calF(0<\gamma < \gammax,t)  &=&  \left(\dfrac{t}{\tini} \right)^4
\left(\dfrac{\aini}{a}\right)^3 \tini^4\,
\calNini\negthickspace\left\{\left[\gamma + \gammad\left(1 -
      \dfrac{\tini}{t} \right) \right] t \right\}\nonumber \\ && +   \Cc (\gamma +
\gammad)^{2 \chic - 3} \funcc{\dfrac{\gammad}{\gamma + \gammad}} \nonumber \\ && -
\Cc (\gamma + \gammad)^{2\chic - 3} \left(\dfrac{t}{\tini}
\right)^{2\chic+1} \left(\dfrac{\aini}{a} \right)^3
\funcc{\dfrac{\gammad}{\gamma + \gammad} \dfrac{\tini}{t}},
\end{eqnarray}
where we have defined
\begin{equation}
\label{eq:gammax}
\gammax(t) \equiv (\gammac + \gammad) \dfrac{\tini}{t} - \gammad.
\end{equation}
Let us stress that, although the solution is defined on three
different domains, it is continuous by construction. The ``middle
domain'' $\gamma\in[\gammax,\gammac]$ owes its existence to two
effects. Loops with a size $\ell < \gammac t$ at time $t$ can either
come from ``on-site production'' with a $\gamma<\gammac$, or they may
have started out earlier as larger loops, having $\gamma>\gammac$,
which then have shrunk into the domain $\gamma < \gammac$ by
gravitational radiation. Of course, the same mechanism applies to all
loops in the initial distribution, as it can be seen from the argument
of $\calNini$ in Eq.~(\ref{eq:finalsol}), weighted by a dilution
factor $a^{-3}$. The third domain $\gamma < \gammax$ corresponds to
the virgin population of loops that started out with a $\gamma <
\gammac$ and which knows nothing about shrunk loops produced at
$\gamma > \gammac$. As one expects, these ``non-contaminated'' loops
cannot exist indefinitely and this domain of the solution only exists
transiently. From Eq.~(\ref{eq:gammax}), we find that $\gammax(\tx) =
0$ for
\begin{equation}
  \dfrac{\tx - \tini}{\tini} = \dfrac{\gammac}{\gammad}\,.
\end{equation}
Clearly, if $\gammac \ll \gammad$, this region exists only for a
negligible duration after $\tini$. Even in the unlikely case of
$\gammac = \gammad$, one finds $\tx = 2 \tini$, i.e. at most one
Hubble time. In the following, we neglect this $\gamma$ domain,
i.e. we shall use only the first two solutions given in
Eq.~(\ref{eq:finalsol}) for the loop number density distribution.

\subsection{Asymptotic expansions}

We are now in a position to determine the unknown constants in
Eq.~(\ref{eq:finalsol}) by matching them to the scaling solutions
obtained in NG numerical simulations~\cite{Ringeval:2005kr}. Since
$\calNini(\ell> {\horizon}_{|\uini}) = 0$, for any fixed $\gamma$
there exists a time at which all terms involving the initial loop
distribution $\calNini$ in Eq.~(\ref{eq:finalsol}) vanish. This
relaxation from the initial loop distribution towards the scaling
regime will be discussed in detail in Sec.~\ref{sec:cosmo}. Taking the
limit $t \gg \tini$ and restricting our attention to the domain probed
by numerical simulations, i.e. $\gamma \gg \gammad$, one gets
\begin{equation}\label{eq:pllarge}
  t^4 \calF(\gamma \gg \gammad, t \gg \tini) \simeq C \,\gamma^{2 \chi -3}.
\end{equation}
This expression matches with Eq.~(\ref{eq:scaling}) provided that
\begin{equation}
\label{eq:fitconsts}
C = \const (1-\nu)^{3-p}, \qquad \chi = 1 - \dfrac{p}{2}\,.
\end{equation}
As discussed in the introduction, the PR functional form of the loop
production function therefore yields a scaling loop distribution which
matches with the one observed in the numerical simulations. Fixing the
value of the above coefficients to those given in
Eq.~(\ref{eq:power-law}) can be considered as a renormalisation
procedure to include the additional sources of loop formation, such as
fragmentation and reconnection, not considered in the genuine
analytical PR model.

\begin{figure}
\begin{center}
\includegraphics[width=0.7\textwidth]{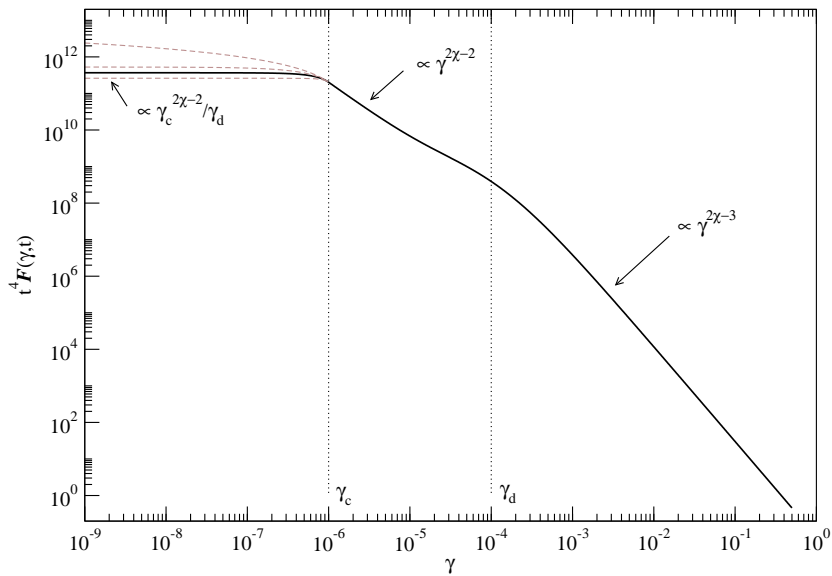}
\caption{The loop number density distribution $t^4 \calF(\gamma,t)$ in
  the full scaling regime, i.e. when all terms involving $\tini$ in
  Eq.~(\ref{eq:finalsol}) can be neglected. The solution has been
  plotted for the radiation era ($\nu=1/2$) with $\chi$ and $C$ given
  by Eq.~(\ref{eq:fitconsts}) and $\chic=2$. The decay and
  gravitational backreaction scales have been fixed to the arbitrary
  values $\gammad=10^{-4}$ and $\gammac=10^{-6}$, respectively. The
  effects of changing $\chic$ are illustrated by the dashed deviations
  which appear only for $\gamma < \gammac$. From bottom to top, they
  correspond to the choices $\chic=4$, $\chic=1.5$ and the logarithmic
  divergent case $\chic=1$.}
\label{fig:scalrad}
\end{center}
\end{figure}

In order to gain some intuition about the solution given in
Eq.~(\ref{eq:finalsol}), we also determine its asymptotic forms in the
two following limits. On the one hand, at length scales smaller than
$\gammad$, but larger than $\gammac$ we recover the result of
Ref.~\cite{Rocha:2007ni},
\begin{equation}
\label{eq:plmedium}
  t^4 \calF(\gammac < \gamma \ll \gammad, t \gg \tini) \simeq \dfrac{C
    \mu}{2 -2\chi} \dfrac{\gamma^{2\chi-2}}{\gammad}\,.
\end{equation}
Hence, in this domain the loop number density distribution is in
scaling but with a power law exponent reduced by one. On the other
hand, for loops smaller than the gravitational backreaction scale
$\gammac$, we obtain for $\chic > 1$ that
\begin{eqnarray}
\label{eq:scaleinv}
t^4 \calF(\gamma \ll \gammac, t \gg \tini)  &\simeq & K
\left(\dfrac{\gammac + \gammad}{\gammad} \right)^{4-3 \nu}  \nonumber \\ && +
\Cc \gammad^{2\chic-3} \dfrac{\Gamma(\muc+1)
  \Gamma(2\chic-2)}{\Gamma(2\chic -2 + \muc)}\,,
\end{eqnarray}
which does no longer depend on $\gamma$. As a result, the loop number
distribution under the gravitational backreaction scale also reaches a
scaling regime and even becomes scale independent. Let us further
mention the extreme case $\chic=1$, which, as previously discussed,
yields a logarithmic divergence (in the radiation era, the matter
epoch will be discussed in the next section),
\begin{equation}\label{eq:pldiv}
  t^4 \calF(\gamma \ll \gammac,t\gg \tini) \simeq \dfrac{\Cc
    \muc}{\gammad} \ln\left(\dfrac{\gammad}{\gamma} \right).
\end{equation}
The constant value of Eq.~(\ref{eq:scaleinv}) can be further expanded
in the limit $\gammac \ll \gammad$. Using Eqs.~(\ref{eq:constsc})
and (\ref{eq:konst}), one finally finds that 
\begin{equation}
\label{eq:pltiny}
 t^4 \calF(\gamma \ll \gammac \ll \gammad, t \gg \tini) \simeq
 \dfrac{C \mu}{2 - 2\chi} \dfrac{\gammac^{2\chi-2}}{\gammad}\,,
\end{equation}
which does no longer depend on $\chic$. Note that
Eq.~(\ref{eq:pltiny}) smoothly connects onto Eq.~(\ref{eq:plmedium})
at $\gamma=\gammac$, as it is to be expected. The loop number density
distribution is sensitive only to the scale at which gravitational
backreaction occurs (at $\gammac$), but not how it occurs. This is
fortunate since the precise form of the loop production function below
the backreaction scale $\gammac$ in unknown. For cosmological
purposes, the only quantity that one needs to determine is thus
$\gammac$. By comparing Eqs.~(\ref{eq:plmedium}) and (\ref{eq:pltiny}),
the physical interpretation is clear: the loop production function at
$\gamma< \gammac$ is always \emph{much} smaller than for
$\gamma>\gammac$ (apart in the extreme case $\chic=1$). As a result,
most loops of size $\gamma<\gammac$ at any given time were originally
produced with a $\gamma>\gammac$ and then shrunk down into the
$\gamma<\gammac$ domain by gravitational radiation. When $\gammac$ is
not much smaller than $\gammad$, then $\chic$ encodes how smooth the
change of behaviour from Eq.~(\ref{eq:plmedium}) to
Eq.~(\ref{eq:pltiny}) is, but still does not affect the value of $t^4
\calF(\gamma=0,t)$ significantly (see Fig.~\ref{fig:scalrad}).

In Fig.~\ref{fig:scalrad}, we have plotted $t^4 \calF(\gamma,t)$ in
the radiation era and in the limit $t \gg \tini$, i.e. when neglecting
all terms in Eq.~(\ref{eq:finalsol}) involving a dependence on the
initial time $\tini$. One recovers the three above mentioned domains
with their power law behaviours of Eqs.~(\ref{eq:pllarge}),
(\ref{eq:plmedium}) and (\ref{eq:pltiny}). Note that the overall shape
does no longer depend on cosmic time $t$ but is universal: the loop
distribution function is in its so-called ``full scaling'' regime. The
black curve has been obtained with $\chic=2$ whereas the dashed gray
curves correspond, from bottom to top, to $\chic=4$, $\chic=1.5$ and
$\chic=1$, respectively. As expected from Eq.~(\ref{eq:pltiny}), apart
the logarithmic divergence when $\chic=1$, deviations due to the
detailed shape of the loop production function below the gravitational
backreaction scale remain negligible.

Similar limits can be derived in the matter era by choosing $C$ and
$\chi$ such that they match the relevant numbers of
Eq.~(\ref{eq:power-law}). The resulting loop distribution has the same
overall shape as Fig.~\ref{fig:scalrad}, with however different values
for the power law indices. Notice also that the values of $\gammac$ in
the radiation and matter era do not need to be the same.

In the next Section, we use the full solution given in
Eq.~(\ref{eq:finalsol}) to discuss how the loop number density
distribution relaxes towards its attractors in both the radiation and
the matter era.

\section{Relaxation towards scaling}
\label{sec:cosmo}

In this Section, we first discuss the relaxation of the loop number
density from a given initial distribution $\calNini(\ell)$ in the
radiation era. The motivation for these calculations comes from the
fact that just after the string formation time, the loop number
density is expected to be different from its cosmological
attractor. Therefore, we want to determine how long it takes to relax
towards the universal form of Fig.~\ref{fig:scalrad}. As we show
below, for any reasonable initial distribution and string formation
temperature, $\calF(\gamma,t)$ is expected to be in full scaling well
before BBN. The second problem concerns the relaxation of the
radiation scaling loop distribution towards the matter era
attractor. Put it in a different way, the question that one may ask is
whether we can still observe traces of loops formed in the radiation
era in the matter loop distribution. Our objective here is to express
our results directly in units of interest in our observable
Universe. For this reason, we express the cosmic time $t$ in terms of
the redshift $z$ according to Eqs.~(\ref{eq:trad}) and
(\ref{eq:tmat}). We moreover use the currently favoured values of the
cosmological parameters $h=0.72$, $\OmegaM h^2=0.13$ and $\OmegaR
h^2=2.471\times 10^{-5}$~\cite{Komatsu:2010fb}, and all lengths and
times are expressed in units of $\Mpc$ unless specified
otherwise. Note that for these cosmological parameters we find from
the definition of $z_{*}$ in Eq.~(\ref{eq:zcross}) that
$z_{*}\simeq3059$.

\subsection{Radiation era}
\label{sec:radera}
As a well-motivated example, let us assume that the initial loop number
density distribution is given by the Vachaspati--Vilenkin (VV) random
walk model of Ref.~\cite{Vachaspati:1984},
\begin{equation}
\label{eq:inivv}
\tini^4 \calNini(\ell) = \Ci \left(\dfrac{\tini}{\ell}\right)^{5/2}.
\end{equation}
Here, the constant $\Ci$ depends on the initial correlation length
$\xi$ of the string network, an order of magnitude estimate being
given by~\cite{Vachaspati:1984}
\begin{equation}
  \Ci \simeq \left(\dfrac{\tini}{\xi} \right)^{3/2}.
\end{equation}
From Kibble's argument, causality imposes that $\xi/\horizon(\tini)
\le 1$ and $\Ci \ge 1$. A maximum value for $\Ci$ can be estimated by
imposing that the energy density of the string network at formation
saturates the radiation energy density, i.e.
\begin{equation}
  H^2(\tini) \simeq \dfrac{8 \pi G}{3}  \dfrac{2 U}{\xi^2}\,.
\end{equation}
Neglecting constant coefficients, one then gets
\begin{equation}
\left(\dfrac{\xi}{\tini}\right)^2 \simeq GU,
\end{equation}
such that
\begin{equation}
\label{eq:cibounds}
1 \le \Ci \le (GU)^{-3/4}.
\end{equation}
The loop distribution of Eq.~(\ref{eq:inivv}) makes sense only for
$\ell \ge \xi$, otherwise no loop can be produced. Also $\ell \le
{\horizon}_{|\uini}$, with ${\horizon}_{|\uini} = 2 \tini$ in the radiation
era (since the long strings are already included in the loop
production function, see the discussion above), and $\calNini(\ell > 2
\tini) = 0$. The maximum value of $t^4 \calNini$ then occurs for
$\ell=\xi$ and can be estimated by
\begin{equation}
\tini^4 \calNini(\xi) \simeq \Ci^{8/3} \le (GU)^{-2}.
\end{equation}
By examining Eq.~(\ref{eq:finalsol}), let us first remark that since
$\calNini$ is maximal at a fixed physical length $\ell=\xi$, any
effect associated with the initial loop distribution should be peaked
at a non-constant $\gamma_\xi(t) =\xi/t$ and then travel towards
smaller and smaller $\gamma$ values. This is what is observed in
numerical simulations during relaxation of the initial conditions
towards the loop scaling regime~\cite{Ringeval:2005kr}. The above
derivation does not consider the damped evolution epoch that is
supposed to take place just after the phase transition responsible for
string formation~\cite{Vilenkin:2000}. During this period, the
background radiation is expected to smooth strings and loops such that
some of the small loops should disappear. The above estimate should
therefore be viewed as a upper bound for $\calNini$.

Using the VV initial distribution of Eq.~(\ref{eq:inivv}), it is
immediate to check that, at fixed $\gamma$, the time dependence of all
terms involving $\calNini$ in Eq.~(\ref{eq:finalsol}) cancels. In
other words, the initial loop density distribution remains, during
some amount of time, on an equal footing with the freshly
produced loops. However, since $\calNini(\ell>2\tini)=0$, this regime can
only be transient since the vanishing initial distribution at super-horizon 
loop sizes will ``sweep leftwards'' towards smaller $\gamma$ as time 
goes on. For any fixed value of $\gamma$, all terms
involving $\calNini$ in Eq.~(\ref{eq:finalsol}) vanish after a time
$\thor$ given by
\begin{equation}
\label{eq:tv}
\dfrac{\thor(\gamma)}{\tini} = \dfrac{2 + \gammad}{\gamma + \gammad}\,.
\end{equation}
The relaxation time $\thor/\tini$ is therefore a function of the scale
of interest $\gamma$. For $\gamma \gg \gammad$ (or $\gammad=0$), it
varies as $2/\gamma$ such that it takes more time for the smaller
loops to reach the scaling attractor, another property observed in the
numerical simulations of Ref.~\cite{Ringeval:2005kr}. For the smallest
values $\gamma \ll \gammad$, the relaxation time saturates to
$\thor/\tini \simeq 2 /\gammad$, which is completely determined by the
gravitational decay length scale. In terms of redshift,
Eq.~(\ref{eq:tv}) reads
\begin{equation}
\label{eq:zv}
\dfrac{1+\zh(\gamma)}{1+\zini} = \sqrt{\dfrac{\gamma + \gammad}{2 + \gammad}}\,,
\end{equation}
which is limited from below by 
\begin{equation}
\dfrac{\zh}{\zini} > \sqrt{\gammad/2}~.
\label{eq:zv2}
\end{equation}
Since $\gammad$ can be small, it may take quite a long redshift range
for the initial loop distribution to vanish and the observability of
such transient effects also depends on $\zini$.

If one assumes that the correlation length $\xi$ is approximately $\xi
\simeq 1/\sqrt{U}$, then Eq.~(\ref{eq:cibounds}) implies that $\mpl
\tini$ can vary from $1/\sqrt{GU}$ to $1/(GU)$, with $\mpl$ denoting
the Planck mass. Taking the latter value and using Eq.~(\ref{eq:trad})
gives the lower bound
\begin{equation}
  \zini > \dfrac{1}{\OmegaR^{1/4}} \sqrt{\dfrac{\mpl}{H_0}} \sqrt{GU}
  \simeq 10^{31}\sqrt{GU}.
\label{eq:zini_lower}
\end{equation}
Combined with Eqs.~(\ref{eq:gammagu}) and (\ref{eq:zv2}), one gets
\begin{equation}
  \zh > \dfrac{ GU }{\OmegaR^{1/4}} \sqrt{\dfrac{\Gamma \mpl}{H_0}}
  \gtrsim 10^{32} GU.
\end{equation}
For string tensions of current cosmological interest, $GU \gtrsim
10^{-10}$ such that $\zh \gtrsim 10^{22}$: the decay of the initial
loops is over well before BBN ($z_{\mathrm{nuc}} \simeq 10^{10}$). In
fact, for all string tensions such that $GU > 10^{-28}$ ($\sqrt{U} >
100\, \TeV$), it is safe to claim that any initial loop remnants have
been radiated away at the time of equality between radiation and
matter ($\zeq\approx10^{4}$). Using the lowest value $\tini=\xi$, the
same statement would apply for strings as light as $GU \simeq
10^{-37}$, i.e. $\sqrt{U} \simeq 1\,\GeV$.

\begin{figure}
\begin{center}
  \includegraphics[width=0.7\textwidth]{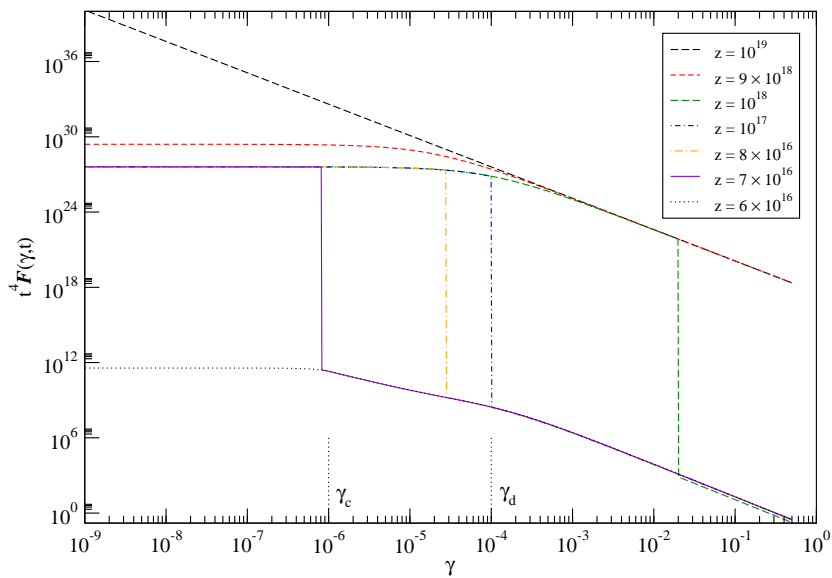}
  \caption{Relaxation towards scaling of the loop number density in
    the radiation era from a Vachaspati--Vilenkin configuration at
    $\zini=10^{19}$. The step, whose position sweeps leftwards over
    the entire range of $\gamma$ with decreasing redshift, corresponds
    to the distance to the horizon at the string formation time above
    which $\calNini=0$. It disappears after a finite time given by
    Eq.~(\ref{eq:tv}) such that it is no longer visible at $z=6 \times
    10^{16}$ and the loop number density distribution is on the
    cosmological attractor.}
\label{fig:relaxrad}
\end{center}
\end{figure}

We have just shown that all terms involving $\calNini$ in
Eq.~(\ref{eq:finalsol}) can be reasonably neglected at all redshifts
$z<\zh$. All the other terms involving $\tini$ exhibit a power law
decrease, up to a correction factor by the hypergeometric function
which rapidly equals unity. In the radiation era, the biggest terms in
Eq.~(\ref{eq:finalsol}) are those appearing in the expression for
$\gamma > \gammac$ and evolve as
\begin{equation}
  \left(\dfrac{t}{\tini} \right)^{2\chi+1} \left(\dfrac{\aini}{a}\right)^3 
  \simeq  \left(\dfrac{z}{\zini} \right)^{1-4\chi}.
\end{equation}
Using the value $1-4\chi \simeq 0.2$ from Eq.~(\ref{eq:power-law}),
they  contribute less than $10\%$ as soon as $z/\zini < 10^{-5}$,
i.e. for
\begin{equation}
\label{eq:zpercent}
z< \zp \equiv 10^{26} \sqrt{GU},
\end{equation}
using Eq.~(\ref{eq:zini_lower}). Again, for all relevant values of
$GU$, these terms cancel well before BBN.

We have plotted in Fig.~\ref{fig:relaxrad} the numerical solution
stemming from Eq.~(\ref{eq:finalsol}) for an initial loop distribution
given by Eq.~(\ref{eq:inivv}). For illustrative purposes, we have
chosen $GU\simeq 10^{-14}$ such that the string formation redshift is
at least $\zini \simeq 10^{19}$ while letting the gravitational effect
at the same value as before. The relaxation towards scaling is
completed at $z \simeq 10^{16}$, as one expects from
Eq.~(\ref{eq:zpercent}).

\subsection{Matter era}

It follows from the previous discussion that the cosmic string loop
number density distribution expected at the end of the radiation era
is given by its scaling solution described by the power laws of
Eqs.~(\ref{eq:pllarge}), (\ref{eq:plmedium}) and
(\ref{eq:pltiny}). Taking this shape as the initial distribution
$\calN_{*}$ at $z=\zcross$ [see Eq.~(\ref{eq:zcross})] for the
subsequent evolution in the matter era, we can now discuss how the
loop number density relaxes towards its new scaling solution in the
matter era.

For $\nu=2/3$, it follows from Eq.~(\ref{eq:defmu}) that $\mu=1-2\chi$
and the hypergeometric function in Eqs.~(\ref{eq:deffunc}) simplifies
to the polynomial expression~\cite{Gradshteyn:1965aa}
\begin{equation}
\label{eq:funcmat}
  \func{}{x} = \dfrac{1}{(1-x)^{2-2 \chi}} \left(1 - \dfrac{x}{2 -
    2\chi} \right),
\end{equation}
and similarly for $\func{c}{x}$, provided $\chic \neq 1$. In the
matter era, the extreme case $\chic=1$ would produce a logarithmic
divergence in time for $\gamma < \gammac$ suggesting that it is no
longer physically acceptable. Therefore, this extreme case will not be
considered in the following. Concerning the values of $\gammad$ and
$\gammac$, one does not expect the former to change significantly from
the radiation to the matter era [as one can see from
Eq.~(\ref{eq:gammagu})].  However, given that $\gammac$ is closely
related to the tangent correlator [see Eq.~(\ref{eq:gammac})], it
assumes a lower value in the matter dominated epoch.

As for the radiation era, it suffices to discuss the relaxation of the
loop number density distribution in the regime $\gamma > \gammac$
since the smaller structures relax even faster. We denote by
$\chirad$, $\Crad$ and $\chimat$, $\Cmat$ the respective values of
$\chi$ and $C$ in the radiation and matter era. The time dependence
associated with the terms involving $\calN_{*}$ (the matter era
equivalent of $\calNini$ for the radiation epoch) in
Eq.~(\ref{eq:finalsol}) shows that, unlike when starting from the VV
distribution, they are now damped during matter domination and become
sub-dominant compared to the scaling solution as soon as
$z<\zd(\gamma)$, with
\begin{equation}
  \dfrac{1+\zd(\gamma)}{1+\zcross} = \left(\dfrac{\Cmat}{\Crad}\right)^{1/(3/2 - 3\chirad)}
  \gamma^{2(\chimat-\chirad)/(3/2 - 3\chirad)}.
\end{equation}
This expression was derived assuming $\gamma \gg \gammad$. Using
the numerical values of Eq.~(\ref{eq:power-law}), this corresponds to
\begin{equation}
\label{eq:zdamping}
\dfrac{1+\zd(\gamma)}{1+\zcross} \simeq 0.16 \, \gamma^{0.2}.
\end{equation}
The damping terms hence are more efficient for the larger length
scales and the scaling solution takes over after an expansion factor
of typically $\mathcal{O}(10)$, i.e. when $\zd \simeq 300$.  For
$\gammac < \gamma \ll \gammad$, one finds the same expression as in
Eq.~(\ref{eq:zdamping}) up to the replacement $\gamma \rightarrow
\gammad$. As a result, if $\gammad$ is very small, the scaling
solution for the small loops may not overcome $\calN_{*}$ before
today. For instance, if $\gamma \le \gammad = 10^{-14}$, one finds
$\zd \simeq 0$. Physically, it means that some of the small loops
today may actually come from the radiation era. Notice that, in
addition to the damping effect, there is the horizon cutoff
$\calN_{*}(\ell > 2 \tcross) = 0$. This imposes that all terms
involving $\calN_{*}$ exactly vanish for $z<\zh(\gamma)$ with
\begin{equation}
\label{eq:zhmat}
\dfrac{1+\zh(\gamma)}{1+\zcross} = \left(\dfrac{\gamma + \gammad}
  {2 + \gammad} \right)^{2/3}.
\end{equation}
However, compared to Eq.~(\ref{eq:zdamping}), $\zh < \zd$ such that
the horizon cutoff is less efficient than damping. Let us remark that
for the typical value of $GU \simeq 10^{-7}$, Eq.~(\ref{eq:gammagu})
implies that $\gammad \simeq 10^{-5}$ and the residual loops from the
radiation era disappear around $\zh \simeq 30$.  Finally, the second 
term involving a dependence on $t_{*}$, the initial time 
for the matter epoch, in Eq.~(\ref{eq:finalsol}) evolves as
\begin{equation}
  \left(\dfrac{t}{t_{*}} \right)^{2\chimat+1} \left(\dfrac{a_{*}}{a} \right)^3
  \simeq \left(\dfrac{1+z}{1+\zcross}\right)^{3/2-3\chimat},
\end{equation}
and contributes less than $10\%$ as soon as $z<\zp \simeq 65$,
independently of the value of $\gammad$.
\begin{figure}
\begin{center}
  \includegraphics[width=0.7\textwidth]{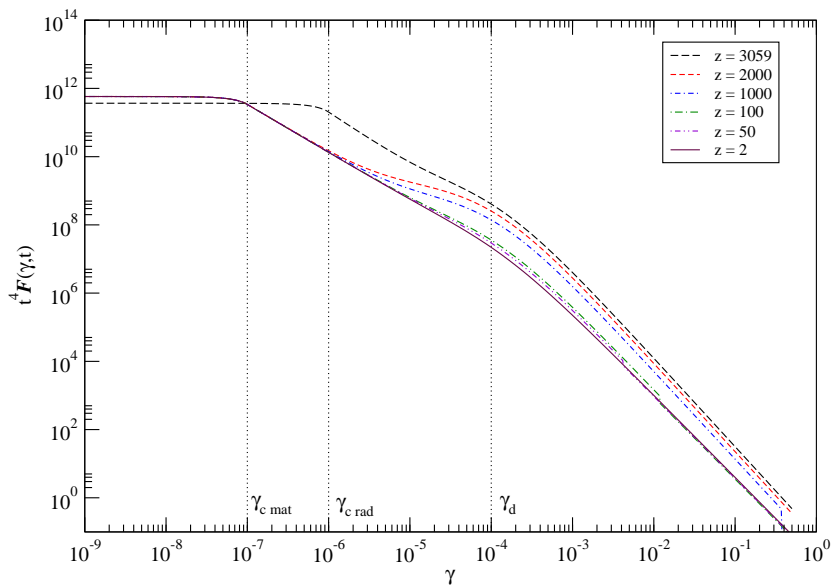}
  \caption{Relaxation of the loop number density distribution towards
    the matter era scaling from the radiation era scaling solution at
    $z=\zcross$. We have chosen the arbitrary values
    $\gammad=10^{-4}$, ${\gammac}_{|\urad} = 10^{-6}$ and
    ${\gammac}_{|\umat}=10^{-7}$ while the other constants have been
    fixed to match numerical simulations [see
    Eq.~(\ref{eq:power-law})]. The relaxation takes longer for $\gamma
    \simeq \gammad$ and is completed at $z\simeq 70$. Lower values of
    $\gammad$ would increase the relaxation time according to
    Eq.~(\ref{eq:zdamping}) such that some small residual loops may
    still be present today.}
\label{fig:relaxmat}
\end{center}
\end{figure}
In Fig.~\ref{fig:relaxmat}, we have plotted the loop distribution in
the matter era when $\calN_{*}$ is given by the radiation era scaling
solution. As expected from the above discussion, we find the
relaxation time to be longer for loops with size $\gammac < \gamma
\lesssim \gammad$ while the redshift at which the matter scaling
solution dominates is compatible with Eq.~(\ref{eq:zdamping}). 

To conclude this section, the loop number density distribution today
is expected to be in full scaling on all length scales provided
$\gammad > 10^{-14}$. If $\gammad < 10^{-14}$, then some of the loops
having a size $\gammac < \gamma \lesssim \gammad$ have been formed in
the radiation era and the distribution is accordingly modified. On the
other length scales, the distribution matches with the matter era
attractor. For those cases, one can use the full expression given in
Eq.~(\ref{eq:finalsol}) to get the precise loop number density
distribution, as plotted in Fig.~\ref{fig:relaxmat}.

\subsection{Cosmological applications}

\subsubsection{Density parameter of cosmic string loops}

We may now use our knowledge of the loop number density distribution
in the matter era to compute various quantities describing the
cosmological impact of these loops today. For simplicity, we neglect
the recent change in the universal expansion from matter to vacuum
energy domination and continue to use Eq.~(\ref{eq:tmat}) to relate
cosmic time to the redshift. Furthermore, we assume that
$\gammad>10^{-14}$ such that the distribution is in its (matter era)
full scaling regime.  From Eqs.~(\ref{eq:finalsol}) and
(\ref{eq:funcmat}), the loop number density distribution then
simplifies to
\begin{eqnarray}
\label{eq:calFmat}
  t^4 \calF(\gamma\geq\gammac,t) & = & \dfrac{C}{\gamma^{2-2\chi}
    (\gamma + \gammad)} \left(1- \dfrac{1}{2-2\chi}
  \dfrac{\gammad}{\gamma + \gammad} \right), \nonumber \\ 
t^4 \calF(\gamma <
\gammac,t) & = & \dfrac{\Cc \gamma^{2\chic -2}}{\gamma + \gammad}
\left(1 + \dfrac{1}{2\chic -2} \dfrac{\gammad}{\gamma + \gammad}
\right)  + \Konst \left(\dfrac{\gammac + \gammad}{\gamma +
    \gammad} \right)^2,
\end{eqnarray}
where $\Cc$ and $\Konst$ are explicitly given in
Eqs.~(\ref{eq:constsc}) and (\ref{eq:konst}) by using
Eq.~(\ref{eq:funcmat}), and we no longer distinguish between $\chirad$
and $\chimat$ as only the latter is relevant for the following
considerations. The energy density in the form of cosmic string loops
of all sizes is given by
\begin{equation}
\label{eq:rholoop}
  \rholoop = \dfrac{U}{t^2} \int_0^3 t^4\calF(\gamma,t) \gamma \, \ud
  \gamma,
\end{equation}
where the upper value comes from the distance to the horizon in the
matter era using $\nu=2/3$ (note that this expression does not include
the Kibble's loops whose contribution remains, however, negligible).
Although Eq.~(\ref{eq:rholoop}) can be explicitly integrated in terms
of hypergeometric functions, the resulting expression is not
particularly illuminating. A very good approximation can be obtained
by piecewise integrating the asymptotic power law expansions of
Eqs.~(\ref{eq:pllarge}), (\ref{eq:plmedium}) and (\ref{eq:pltiny}) in
the three domains $\gamma \leq \gammac$, $\gammac < \gamma \leq
\gammad$, and $\gamma > \gammad$. After some algebra, one finds
\begin{eqnarray}
  \dfrac{\rholoop t^2}{U} &\simeq&  \dfrac{C}{\gammad^{1-2\chi}}
  \Bigg[ \dfrac{1}{4 \chi(1-\chi)(1-2\chi)} - \dfrac{1-2\chi}{4\chi}
    \left( \dfrac{\gammac}{\gammad} \right)^{2\chi}   \nonumber \\ && -
  \dfrac{1}{1-2\chi} \left(\dfrac{\gammad}{3} \right)^{1-2\chi}
  \Bigg],
\end{eqnarray}
which can be further approximated for $\gammac \ll \gammad \ll 1$ by
\begin{equation}
\label{eq:rhonow}
\dfrac{\rholoop t^2}{U} \simeq \dfrac{C}{4 \chi(1-\chi)(1-2\chi)}
\dfrac{1}{\gammad^{1-2\chi}}\,.
\end{equation}
This expression clearly shows that the energy density in the form
of loops is essentially fixed by the gravitational decay scale
$\gammad$ and the scaling function properties through $C$ and
$\chi$. Smaller $\gammad$ implies larger values for
$\rholoop/U$. Notice also that the scaling evolution is similar to
that of the long string ones; the quantity $t^2 \rholoop/U$ is time
independent and universal. Doing the integration exactly and then
expanding the hypergeometric functions gives the same result, up to
the first numerical coefficient which changes by fifteen percents and
reads $C \pi /[(2-2\chi)\sin(2\pi \chi)]$. In terms of the density
parameter $\OmegaO\equiv\rholoop/\rho_{\mathrm{crit}}$, where
$\rho_{\mathrm{crit}}$ is the critical density for a flat universe
today, we find
\begin{equation}
  \OmegaO = \dfrac{3 \pi^2 C}{(1-\chi) \sin(2\pi\chi)} \dfrac{GU}{\gammad^{1-2\chi}}\,,
\end{equation}
which is valid at all scaling times during the matter era. Up to some
small additional dilution factor coming from the recent cosmological
constant domination, this expression may be extrapolated until
today. Using Eq.~(\ref{eq:gammagu}) with $\Gamma \simeq 10^2$ and the
numbers derived before, one finds
\begin{equation}
\OmegaO \simeq 0.10 \times (GU)^{0.59}.
\end{equation}
This cannot exceed $10^{-5}$ with the values of $GU$ currently allowed
by CMB observations, i.e. $GU \le 7\times 10^{-7}$, and hence
Nambu--Goto cosmic string loops cannot provide a viable Dark Matter
candidate, a scenario which has been dicussed in
Ref.~\cite{Cui:2008bd}.

\subsubsection{Number density of loops in a box of size $L$}

Again assuming the full scaling solution in the matter era, we can
estimate the number density of cosmic string loops present around us
today. Taking a box of size $L$, Eq.~(\ref{eq:calFmat}) can be used to
determine the number density $n_L$ of loops having a length $\ell \le
L$ at $t$:
\begin{equation}
t^3 n_L = \int_0^{L/t} t^4 \calF(\gamma,t)\,\ud \gamma.
\end{equation}
Assuming the box is big enough, with $\gammac < \gammad \ll L/t$,
we find the dominant part of the loop density to be independent of $L$,
\begin{equation}
  t^3 n_L  =  \dfrac{2\chic - 2\chi}{2\chic -1} \,
  \dfrac{C}{\gammad \gammac^{1-2\chi}}\,.
\end{equation}
As expected, the dependence in $\chic$ is very weak such that the
first term in the previous equation can be approximated to unity.
Using Eqs.~(\ref{eq:gammagu}) and (\ref{eq:gammac}) with
$\Gamma\simeq10^{2}$ as before and $\Upsilon\simeq10$, yields
\begin{equation}
  n_L \simeq \dfrac{6.1 \times 10^{-5}}{t^3} (GU)^{-1.65}.
\end{equation}
Using $t=t_0$ from Eq.~(\ref{eq:tmat}) evaluated for $z=0$ (which
again amounts to neglecting the recent change to a vacuum energy
dominated universe) with $GU \simeq 7 \times 10^{-7}$ gives the lower
bound
\begin{equation}
n_L \simeq 5.5\times 10^{-6}\,\Mpc^{-3}.
\end{equation}
This number is quite small, but interestingly it scales as a negative
power of $GU$, such that loops from very light cosmic strings are the
more numerous. Our result suggests that lensing
events~\cite{deLaix:1996vc} from loops are more frequent for low
values of $GU$ (but also of smaller amplitude). Note that with the
above string tension, the gravitational backreaction length is at
$\gammac t_0 \simeq 8\,\pc$, and the gravitational decay length at
$\gammad t_0 \simeq 380\,\Kpc$.

\section{Conclusions}\label{sec:conclusions}

In this paper, we have proposed a phenomenological model of
Nambu--Goto cosmic string loop evolution based on the Polchinski--Rocha
loop production function, which we adjusted to fit numerical
simulations.  We have considered both loop decay by gravitational wave
emission and string smoothing due to gravitational backreaction. We
have found that the loop number density distribution rapidly assumes a
universal form on all length scales, which ends up being insensitive
to the details of the backreaction effects. It solely depends on the
scale $\gammac$ below which these effects dominate.  In the matter
era, the energy density parameter of loops $\OmegaO$ is a positive
power of the string tension $GU$ and ends up being completely
negligible. Hence, cosmic string loops are unfit to account for the
Dark Matter contribution to our Universe's energy density. On the
other hand, the density number of loops is boosted by a negative power
of $GU$, meaning that very low tension loops could actually be
abundant in the present Universe.

Our work may be extended along several lines. For instance, since the
universal loop distribution plotted in Fig.~\ref{fig:scalrad} is
significantly different than the loop distributions used so far in the
literature, various constraints on $GU$ coming from gravitational wave
emission, lensing or pulsar-timing should be
re-examined~\cite{Damour:2001bk, Siemens:2006vk, Pshirkov:2009vb,
  Olmez:2010bi, Tuntsov:2010fu}. All along the calculation, we have
kept $\gammad$ and $\gammac$ as undetermined parameters (apart from
the numerical examples) since these parameters may equally well be
used to represent other physical effects. For instance, $\gammad$ does
not need to be given by Eq.~(\ref{eq:gammagu}) but could represent the
decay rate of another energy loss mechanism, such as particle
emission, provided it can still be described phenomenologically by
Eq.~(\ref{eq:gammad}).  Another extension would be the implementation
of a low intercommutation probability (as in the PR model, we have set
this probability $P=1$ throughout this work) to extend our approach to
the evolution of cosmic superstring loops.

\begin{acknowledgments}
  It is a pleasure to thank Patrick Peter and Daniele Steer for
  enlightening discussions. The work of L.~L. and C.~R. is partially
  supported by the Belgian Federal Office for Science, Technical and
  Cultural Affairs, under the Inter-university Attraction Pole Grant
  No. P6/11. The work of M.~S. is partially supported by the Sciences
  \& Technology Facilities Council (STFC--UK), Particle Physics
  Division, under the grant ST/G000476/1 ``Branes, Strings and Defects
  in Cosmology''.
  
\end{acknowledgments}

\section*{References}

\bibliography{bibstrings}
\end{document}